\documentclass[amsmath,amssymb,twocolumn, a4paper,13pt]{revtex4}
\usepackage{amsmath}
\usepackage{amssymb}
\usepackage{amsmath,amsthm,amssymb,amscd}
\usepackage{latexsym}
\usepackage{indentfirst}
\usepackage{subfigure}
\usepackage{graphicx}

\makeatletter
\def\ExtendSymbol#1#2#3#4#5{\ext@arrow 0099{\arrowfill@#1#2#3}{#4}{#5}}
\makeatother

\date{\today}

\newcommand{\atp}{{\textrm{[ATP]} }}

\begin{document}

\title{
\bf Phenomenological analysis of ATP dependence of motor protein

}

\author{Yunxin Zhang}\email[Email: ]{xyz@fudan.edu.cn}
\affiliation{
Laboratory of Mathematics for Nonlinear Science, Centre for Computational System Biology,
School of Mathematical Sciences, Fudan University, Shanghai 200433, China.
}

\begin{abstract}
In this study, through phenomenological comparison of the velocity-force data of processive motor proteins, including conventional kinesin, cytoplasmic dynein and myosin V, we found that, the ratio between motor velocities of two different ATP concentrations is almost invariant for any substall, superstall or negative external loads. Therefore, the velocity of motor can be well approximated by a Michaelis-Menten like formula $V=\atp k(F)L/(\atp +K_M)$, with $L$ the step size, and $k(F)$ the external load $F$ dependent rate of one mechanochemical cycle of motor motion in saturated ATP solution. The difference of Michaelis-Menten constant $K_M$ for substall, superstall and negative external load indicates, the ATP molecule affinity of motor head for these three cases are different, though the expression of $k(F)$ as a function of $F$ might be unchanged for any external load $F$. Verifications of this Michaelis-Menten like formula has also been done by fitting to the recent experimental data.

\end{abstract}

\maketitle

The processive motor proteins, including kinesin, dynein and myosin are essential for biophysical functioning of eukaryotic cells \cite{Howard2001, Vale2003}. Due to the development of experimental instrument \cite{Greenleaf2007, Guydosh2009}, much accurate experimental data have been obtained \cite{Guydosh2009, Mehta1999, Nishiyama2002, Block2003, Uemura2004, Carter2005, Toba2006, Christof2006, Gennerich2007, Chuan2011}. Both conventional kinesin and cytoplasmic dynein move hand-over-hand along microtubules with step size 8.2 nm by converting chemical energy stored in ATP molecules into mechanical works \cite{Schnitzer1997, Hua1997, Asbury2003, Yildiz2004, Toba2006}. Myosin (V or VI) also moves hand-over-hand but along actin filament with step size around 36 nm \cite{Yildiz2003, Uemura2004, Snyder2004, Iwaki2009}. So far, there are many biophysical models to understand the mechanism of motor proteins, including the flashing ratchet model \cite{Astumian1997, Parrondo1998, Christof2006}, Fokker-Planck equation \cite{Risken1989, Lipowsky2000, Zhang20091}. Meanwhile, more detailed mechanochemical models have also been designed to explain the experimental data, and get meaningful biochemical parameters \cite{Rief2000, Fisher2001, Kolomeisky2003, Rosenfeld2004, Gao2006, Zhang2008, Chuan2011}.

In this study, by phenomenological comparison of the velocity-force data of different ATP concentrations, we found that the velocity of processive motor proteins can be described by a Michaelis-Menten like formula $V=\atp k(F)L/(\atp +K_M)$, but might with different constant $K_M$ for substall, superstall and negative external loads. So the limit velocity in saturated ATP solution is $V_*=k(F)L$, and the velocity corresponding to ATP concentration \atp can be obtained simply by multiplying $V_*$ by a constant \atp /(\atp +$K_M$).

\begin{figure}
  \includegraphics[width=270pt]{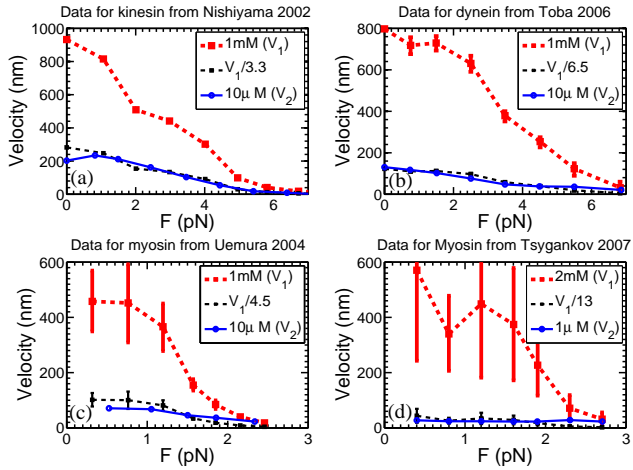}\\
  \caption{For the positive substall external load cases, the velocity $V_2$ of motor protein at low ATP concentration can be well approximated by the velocity $V_1$ at high ATP concentration divided by a constant. (a) For the experimental data of kinesin measured in \cite{Nishiyama2002}, the velocity $V_2$ of \atp=10 $\mu$M can be approximated by $V_1/3.3$ with $V_1$ the velocity of \atp=1 mM. (b) For the data of dynein measured in \cite{Toba2006}, velocity $V_2$ of \atp=10 $\mu$M can be well approximated by $V_1/6.5$ with $V_1$ the velocity of \atp=1 mM. (c) For the data of myosin V measured in \cite{Uemura2004}, velocity $V_2$ of \atp=10 $\mu$M can be well approximated by $V_1/4.5$ with $V_1$ the velocity of \atp=1 mM. (d) For the data of myosin V used in \cite{Tsygankov2007} (derived from \cite{Mehta1999}), velocity $V_2$ of \atp=1 $\mu$M can be well approximated by $V_1/13$ with $V_1$ the velocity of \atp=2 mM.}\label{FigSubstall}
\end{figure}

\begin{figure}
  \includegraphics[width=270pt]{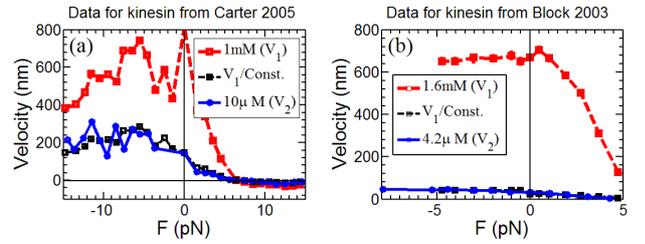}\\
  \caption{For general external load cases, the velocity $V_2$ of kinesin at low ATP concentration can be well approximated by the velocity $V_1$ at high ATP concentration divided by a constant. (a) For the data in \cite{Carter2005}, the velocity $V_2$ of \atp=10 $\mu$M can be well approximated by velocity $V_1$ for \atp=1 mM divided by a constant Const with Const=2.6 for $F<0$, Const=5.6 for $0\le F\le 7$ pN, and Const=2.3 for $F>7$ pN. (b) For the data in \cite{Block2003}, the velocity $V_2$ of \atp=4.2 $\mu$M can be well approximated by velocity $V_1$ for \atp=1.6 mM divided by a constant Const with Const=29 for $F<0$, Const=16 for $F\ge 0$.}\label{FigAllForce}
\end{figure}

\begin{figure}
  \includegraphics[width=120pt]{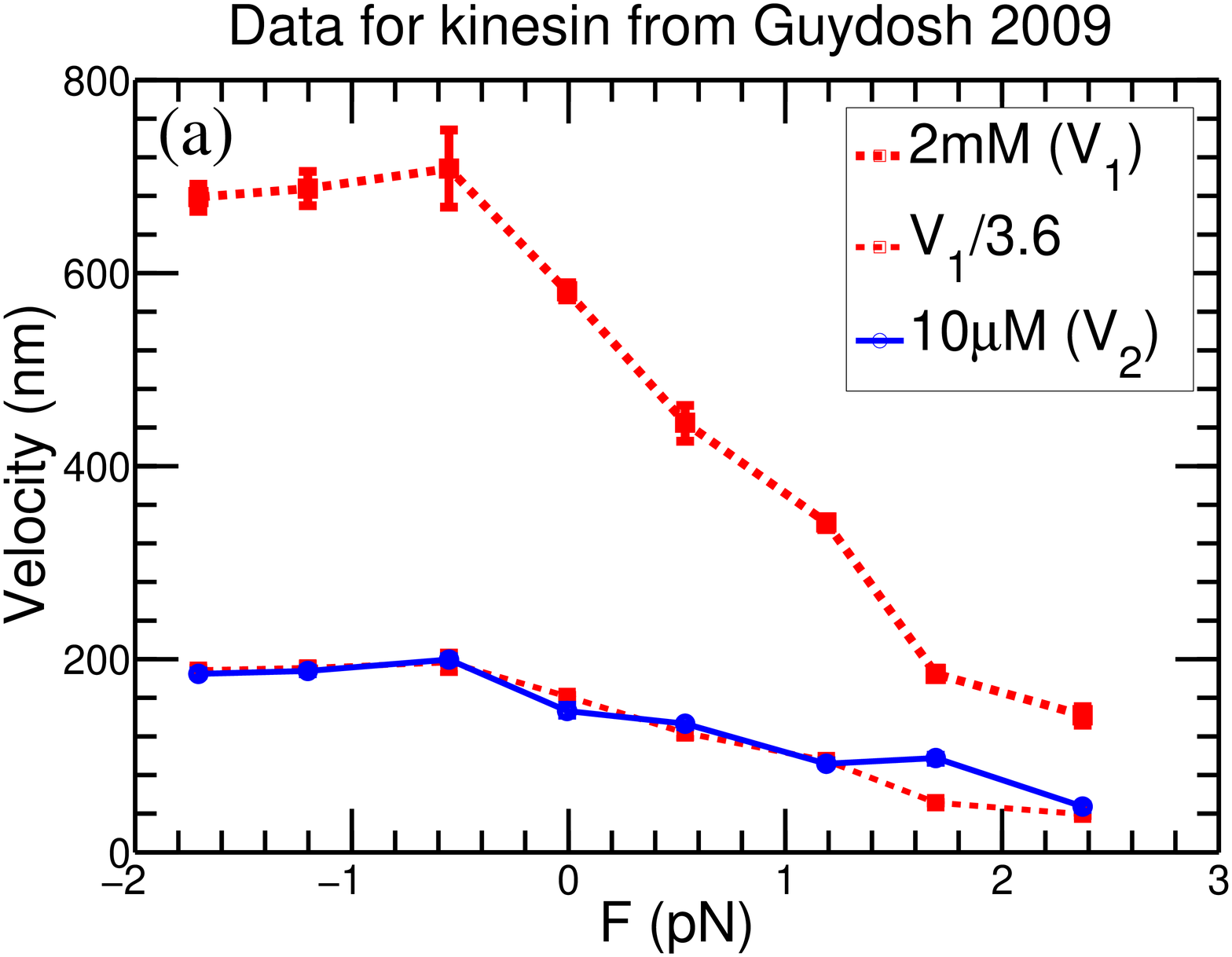}\includegraphics[width=118pt]{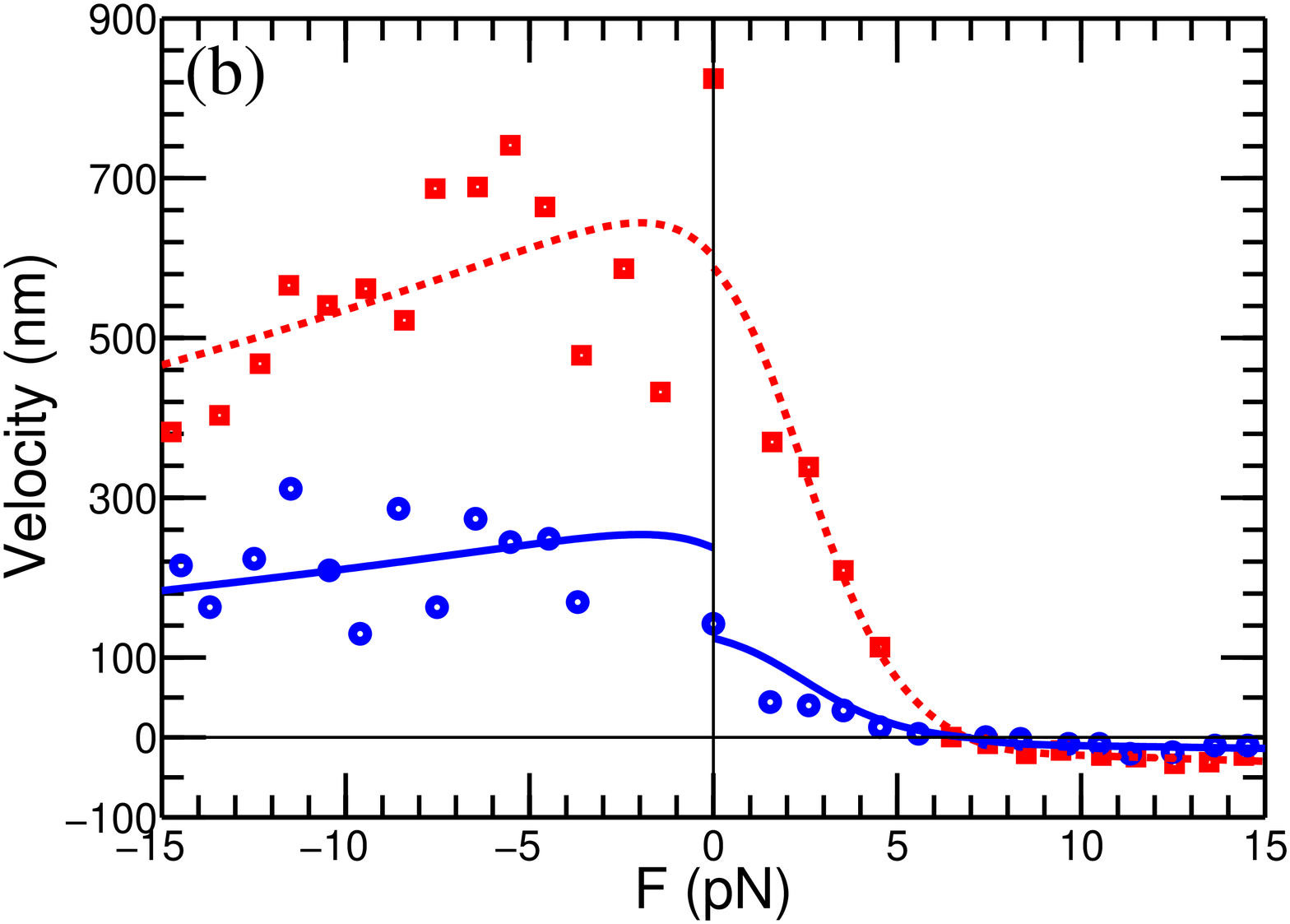}\\
  \caption{(a) For the kinesin data measured in \cite{Guydosh2009}, the velocity $V_2$ of kinesin at low ATP concentration (10 $\mu$M ) can be well approximated by the velocity $V_1$ at high ATP concentration (2 mM) divided by a constant 3.6, which is the same for both substall and negative external load.
  (b)Experimental data for conventional kinesin measured in \cite{Carter2005} and the theoretical prediction using the Michaelis-Menten like formula $V=\atp k(F)L/(\atp +K_M)$. The ATP concentrations are corresponding to \atp=1 mM (dashed line and squares) and 10 $\mu$M (solid line and dots) respectively. The model parameter $K_M$ is 15.8 $\mu$M for $F<0$, 39.2 $\mu$M for $0\le F\le 7$ pN, and 11.9 $\mu$M for $F>7$ pN, others are listed in Tab. \ref{table1}. }\label{FigFitCarter}
\end{figure}

For the sake of comparison, the velocity-force data of kinesin, dynein and myosin are plotted in Figs. \ref{FigSubstall}, \ref{FigAllForce} and \ref{FigFitCarter}(a). In Fig. \ref{FigSubstall}(a), the thick dashed line $V_1$ is the velocity-force data of kinesin for \atp=1 mM obtained by Nishiyama {\it et al} \cite{Nishiyama2002}, and the solid line $V_2$ is for \atp=10 $\mu$M. One can easily see that there is only little difference between the lines $V_2$ and $V_1/3.3$. Similar phenomena can also be found for the velocity-force data of dynein and myosin obtained in \cite{Toba2006, Uemura2004, Tsygankov2007}, see Figs. \ref{FigSubstall}(b,c,d). Generally, these phenomena also exits for negative and superstall force cases, but might with different ratio constant, see Figs. \ref{FigAllForce} and \ref{FigFitCarter}(a) for data of kinesin obtained in Refs. \cite{Guydosh2009, Block2003, Carter2005}. For the kinesin data in \cite{Carter2005}, the ratio constant is about 2.6 for $F<0$, about 5.6 for $0\le F\le 7$ pN, and about 2.3 for $F>7$ pN [see Fig. \ref{FigAllForce}(a)]. For the data in \cite{Block2003}, the ratio constant is about 29 for $F<0$, and about 16 for $0\le F\le 5$ pN [see Fig. \ref{FigAllForce}(b)]. But for the kinesin data measured in \cite{Guydosh2009}, the constant 3.6 works well for both substall and negative external load [see Fig. \ref{FigFitCarter}(a)].

From the above observations about the experimental data plotted in Figs. \ref{FigSubstall} and \ref{FigAllForce}, one can see that the velocity-force relation of motor proteins satisfies $V(F,\atp)=f(\atp)V_*(F)$. Where $V_*=V_*(F)$ is the velocity-force relation at saturated ATP concentration, and obviously $V_*$ can be written as $V_*(F)=k(F)L$ with $L$ the step size of motor protein, and $k(F)$ the force dependent rate to complete one ATP hydrolysis cycle (coupled with one mechanical cycle). The function $f(\atp)$ increases with [ATP], $f(0)=0$ and $f(\atp)=1$ with $\atp\to\infty$. A reasonable form of $f(\atp)$ is $f(\atp)=\atp /(\atp +K_M)$ with a parameter $K_M$ which we called {\it Michaelis-Menten constant}. Finally, the velocity formula can be written as $V(F,\atp)=\atp k(F)L/(\atp +K_M)$.

To verify the above velocity-force formula, the force dependent expression of rate $k(F)$ should be given firstly. Usually, the mechanical coupled cycle of ATP hydrolysis  includes several internal states, here, as demonstrated in the previous mechanochemical model \cite{Fisher2001}, we assume that, in each cycle, there are two internal states, denoted by state 1 and state 2 respectively.
\begin{equation}\label{eq1}
\cdots\rightleftarrows\overbrace{1\rightleftarrows2\rightleftarrows1}^{\textrm{one cycle}}\rightleftarrows\cdots
\end{equation}
Let $u_i, w_i$ be the forward and backward transition rates at state $i$, then the steady state rate $k(F)$ can be obtained as follows \cite{Derrida1983, Fisher2001}
\begin{equation}\label{eq2}
k=\frac{u_1u_2-w_1w_2}{u_1+u_2+w_1+w_2}.
\end{equation}
The force dependence of rates $u_i, w_i$ are assumed to be \cite{Fisher2001}
\begin{equation}\label{eq3}
\begin{array}{ll}
u_1(F)=u_1^0e^{-\theta_1^+FL/k_BT},\quad & u_2(F)=u_2^0e^{-\theta_2^+FL/k_BT},\cr
w_1(F)=w_1^0e^{\theta_1^-FL/k_BT},\quad & w_2(F)=w_2^0e^{\theta_2^-FL/k_BT}.
\end{array}
\end{equation}
Where $k_B$ is the Boltzmann constant, $T$ is the absolute temperature, and $\theta_i^{\pm}$ are {\it load distribution factors} which satisfy $\theta_0^++\theta_1^++\theta_0^-+\theta_1^-=1$. For this two-state model, one can easily get the following formula of motor velocity
\begin{equation}\label{eq4}
\begin{aligned}
V(F,\atp)=&\frac{\atp (u_1u_2-w_1w_2)L}{(\atp +K_M)(u_1+u_2+w_1+w_2)}.
\end{aligned}
\end{equation}

\begin{figure}
  \includegraphics[width=120pt]{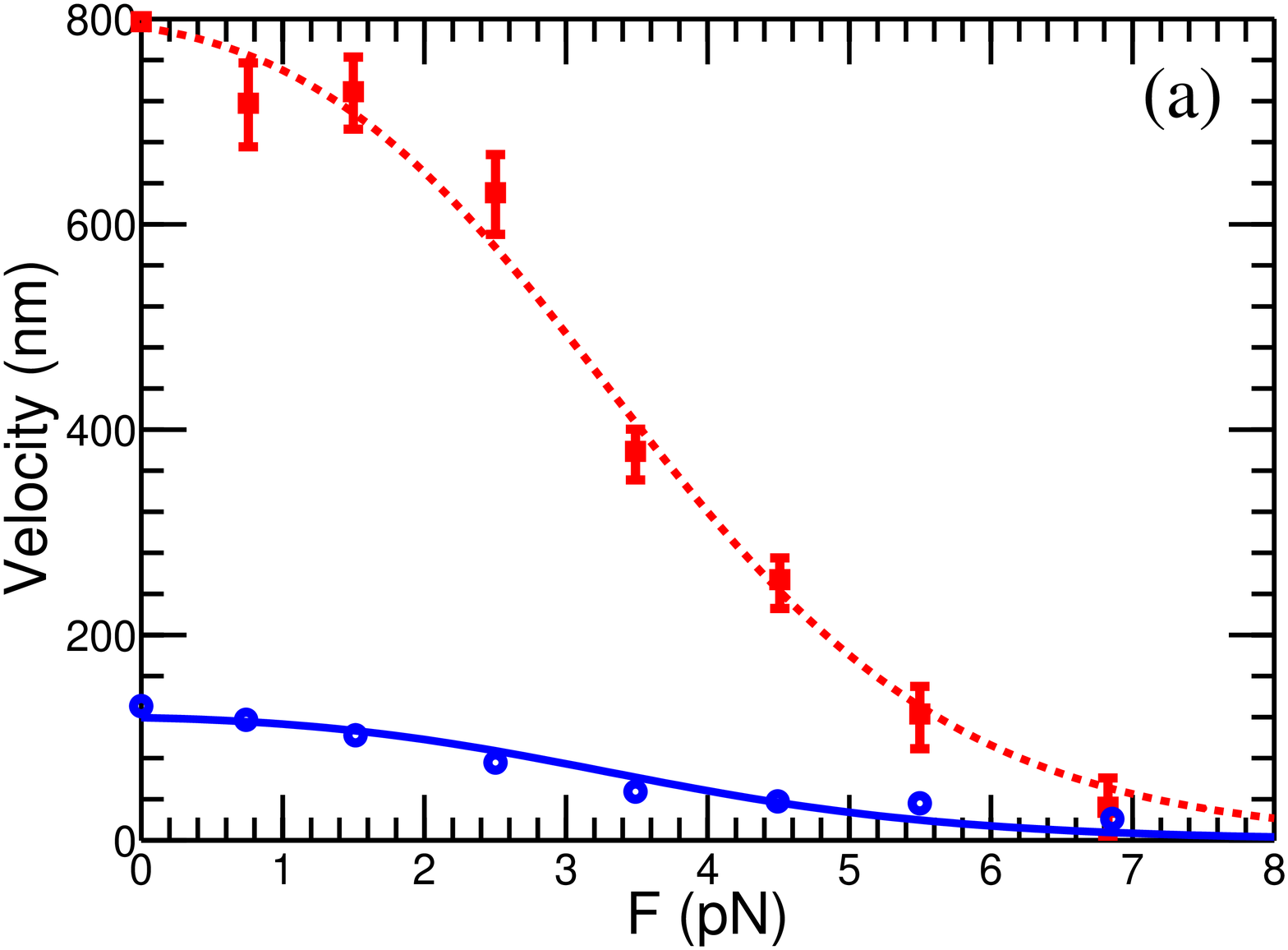}\includegraphics[width=120pt]{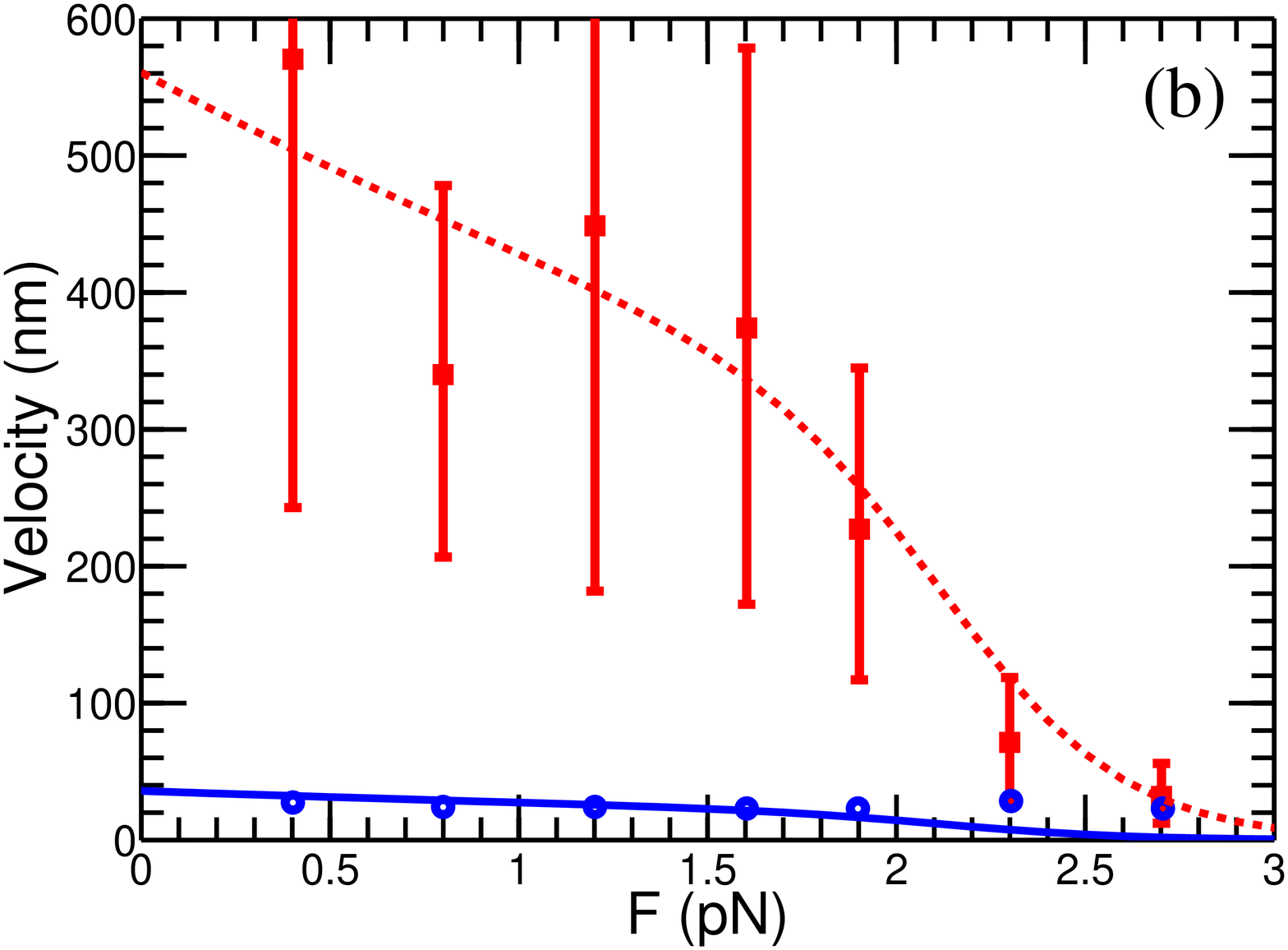}\\
  \caption{Experimental data for cytoplasmic dynein obtained in \cite{Toba2006} and myosin obtained in \cite{Mehta1999} (see also \cite{Tsygankov2007} for the method to get the present values), and the theoretical prediction using the Michaelis-Menten like formula $V=\atp k(F)L/(\atp +K_M)$. (a) The experimental data are for \atp=1 mM (dashed line and squares) and 10 $\mu$M (solid line and dots). The model parameter $K_M=60.3$ $\mu$M for $0\le F\le7$ pN. (b) The experimental data are for \atp=2 mM (dashed line and squares) and 1 $\mu$M (solid line and dots). The model parameter $K_M=14.7$ $\mu$M for $0\le F\le3$ pN.
  }\label{FigFitTobaTsygankov}
\end{figure}

The fitting results of the above velocity-force formula to kinesin data measured in \cite{Carter2005} are plotted in Fig. \ref{FigFitCarter}(b). In which, the Michaelis-Menten constant $K_M=15.8$ $\mu$M for $F<0$, $K_M=39.2$ $\mu$M for $0\le F\le 7$ pN, and $K_M=11.9$ $\mu$M for $F>7$ pN, other parameter values are listed in Tab. \ref{table1}. Meanwhile, the fitting results to the dynein data measured in \cite{Toba2006} and myosin data measured in \cite{Mehta1999} are plotted in Fig. \ref{FigFitTobaTsygankov}(a) and Fig. \ref{FigFitTobaTsygankov}(b) (with Michaelis-Menten constant $K_M=60.3$ $\mu$M and 14.8 $\mu$M) respectively, see also Tab. \ref{table1} for the corresponding parameter values \footnote{The value of $K_M$ obtained in Figs. \ref{FigFitCarter}(b) and \ref{FigFitTobaTsygankov} might not be consistent with the ratio constant used in Figs. \ref{FigSubstall} and \ref{FigAllForce}, since the plots in Figs. \ref{FigSubstall} and \ref{FigAllForce} are just phenomenological illustration, and the ratio constants are obtained by rough estimation. For example, for the dynein data plotted in Fig. \ref{FigFitTobaTsygankov}(a), $K_M=60.3$ $\mu$M means the ratio constant between $V_1$ and $V_2$ is 6.6, but 6.5 is used in Fig. \ref{FigSubstall}(b).}. Note, the step size used in the calculations is  $L=8.2$ nm for motor proteins kinesin and dynein, but $L=36$ nm for myosin V. Certainly, the same fitting process can also be done to other experimental data. The plots in Figs. \ref{FigFitCarter}(b) and \ref{FigFitTobaTsygankov} indicate that, the experimental data of motor proteins can be well reproduced by the Michaelis-Menten like formula (\ref{eq4}), so our phenomenological analysis about the ATP dependence of motor motion is reasonable.

\begin{table}
  \centering
  \caption{Parameter values used in the theoretical predictions of the velocity-force relation for conventional kinesin, cytoplasmic dynein and myosin V: see Figs. \ref{FigFitCarter}(b) and \ref{FigFitTobaTsygankov}(a)(b). The unit of rate $u_i^0, w_i^0$ is $\textrm{s}^{-1}$.}
    \begin{tabular}{c||c|c|c|c||c|c|c|c}
    \hline\hline
       & $u_1^0$ & $u_2^0$  & $w_1^0$  & $w_2^0$ & $\theta_1^+$  & $\theta_1^-$  & $\theta_2^+$  & $\theta_2^-$ \\
    \hline
    kinesin &716.6&4235.5&0.25&13.5&-0.014&0.609&0.378&0.027 \\
    dynein &910.8&$1.15\times10^4$&64.0&0&-0.019&0.019&0.386&0.614 \\
    myosin &584.0&$1.73\times10^4$&2.55&0&0.03&0.43&0.03&0.51 \\
    \hline\hline
  \end{tabular}
  \label{table1}
\end{table}

In summary, in this study, the ATP dependence of motor proteins is phenomenological discussed. Based on the recent experimental data and basic numerical calculations, we found the motor velocity can be well described by a Michaelis-Menten like formula $V=\atp k(F)L/(\atp +K_M)$ with force dependent rate $k(F)$ at saturated ATP. The different values of $K_M$ for substall, superstall and negative external load imply, the ATP molecule affinity of motor head might be different for these three cases, but the basic mechanism in each mechanochemical cycle (either forward or backward) might be the same. An obvious conclusion from our Michaelis-Menten like formula is that the {\it stall force}, under which the mean motor velocity is vanished, is independent of ATP concentration \cite{Carter2005, Toba2006, Gennerich2007}.

\vskip 0.5cm

\acknowledgments{This study is funded by the Natural
Science Foundation of Shanghai (under Grant No. 11ZR1403700).
}

\end{document}